\documentclass[11pt]{article}
\usepackage{blois,epsfig}
\usepackage[latin1]{inputenc}
\usepackage[OT1]{fontenc}

\def\Journal#1#2#3#4{{#1} {\bf #2}, #3 (#4)}

\def\NIMA{{\em Nucl. Instrum. Methods} A}

\def\PRL{\em Phys. Rev. Lett.}
\def\PRC{{\em Phys. Rev.} C}

\def\be{\begin{equation}}
\def\ee{\end{equation}}
\def\bea{\begin{eqnarray}}
\def\eea{\end{eqnarray}}

\begin{document}
\vspace*{4cm}
\title{NEW RESULTS FROM THE NEMO~3 EXPERIMENT}

\author{ V.I. TRETYAK on behalf of the NEMO Collaboration}

\address{Institute Pluridisciplinaire Hubert Curien,\\
 23 Rue du Loess, BP 28, 67037 Strasbourg, France\\
on leave of absence from Joint Institute for Nuclear Research,\\ 
Joliot Curie 6, 41980 Dubna, Russia}

\maketitle\abstracts{
NEMO~3 is a currently running experiment to search for the neutrinoless 
double beta decay and to study the two-neutrino double beta decay with 10kg of 
enriched isotopes. On the basis of the first two years of data taking, a limit 
on the neutrinoless decay  $T_{1/2}^{0\nu} > 5.8\cdot10^{23}$ y at 90\% CL was obtained 
with $^{100}Mo$. The two-neutrino double beta decay half-lives were measured for all 
seven $\beta\beta$ isotopes located in NEMO~3. New results are obtained for two of these
isotopes: $^{48}Ca~T_{1/2}^{2\nu}=[4.4 ^{+0.5}_{-0.4} (stat) \pm 0.4 (syst)]\cdot10^{19}y$ ,
$^{96}Zr~T_{1/2}^{2\nu}=[2.3 \pm 0.2 (stat) \pm 0.3 (syst)]\cdot10^{19}y$.
}

\section{Introduction}
Experimental search for the neutrinoless double beta decay ($0\nu\beta\beta$) is of a major 
importance in particle physics  because if observed, it will reveal the 
Majorana nature of the neutrino ($\nu \equiv \bar{\nu}$) and may allow an 
access to the absolute neutrino mass scale. 
The $0\nu\beta\beta$-decay violates the lepton number and is therefore a direct 
probe for the physics 
beyond the standard model.
A possibility of this process may be related to 
right-handed currents in electroweak interactions,
supersymmetric particles with R-parity nonconservation,
massless Goldstone bosons such as majorons.
In the case of the neutrino-mass mechanism the $0\nu\beta\beta$ decay rate can be written as
\be
 [T_{1/2}^{0\nu}(A,Z)]^{-1} = \langle m_{\nu}\rangle^2 \cdot |M^{0\nu}(A,Z)|^2 \cdot G^{0\nu}(Q_{\beta\beta},Z) , 
\label{eq:nme} 
\ee 
where $\langle m_{\nu}\rangle$ is the effective neutrino mass, $M^{0\nu}$ is the nuclear matrix element  
(NME), and $G^{0\nu}$ is the kinematical factor proportional to the transition energy to the fifth power, 
$Q_{\beta\beta}^5$. 

The spontaneous two-neutrino double beta decay  ($2\nu\beta\beta$)
is a rare second-order weak interaction process. 
The accurate measurement of the $2\nu\beta\beta$-decay is important since 
it constitutes the ultimate background in the search for $0\nu$-decay signal.
It is useful for the test of the nuclear structure and provides valuable input for the 
theoretical calculations of the $0\nu\beta\beta$-decay NME.

The objective of the NEMO~3 experiment is the search for the $0\nu\beta\beta$-decay
and investigation of the $2\nu\beta\beta$-decay with 10 kg of $\beta\beta$ isotopes.
\section{NEMO~3 experiment}
\subsection{The NEMO~3 detector}\label{subsec:detector}
The NEMO~3 has been taking data since 2003 in the Modane 
underground laboratory located in the Frejus tunnel at the depth of 4800 m w.e. 
Its method of $\beta\beta$-decay study is based on the 
detection of the electron tracks in a tracking device and the 
energy measurement in a calorimeter. 

The detector~\cite{nemo3} has a cylindrical shape.
Thin source foils ($\sim 50~mg/cm^2$) are located
in the middle of the tracking volume surrounded by the calorimeter.
Almost 10kg of enriched $\beta\beta$ isotopes (listed in Table~\ref{tab:t12})
were used to produce the source foils.
The tracking chamber contains 6180 open drift cells operating in the Geiger mode.
It provides a vertex resolution of about 1 cm.
The calorimeter consists of 1940 plastic scintillator blocks with photomultiplier readout.
The energy resolution is 14-17\%/$\sqrt{E}$ FWHM. The time resolution of 250 ps 
allows excellect suppression of the crossing electron background.
A 25~G magnetic field is used for charge identification.
The detector is capable of identifying e$^-$, e$^+$, $\gamma$ and $\alpha$ particles
and allows good discrimination between signal and background events.
\subsection{Event selection and background model}
The $\beta\beta$ events are selected by requiring two reconstructed
electron tracks with a curvature corresponding to the negative charge,
originating from a common vertex in the source foil.
The energy of each electron measured in the calorimeter should be higher than 200 keV.
Each track must hit a separate scintillator block.
No extra PMT signal is allowed.
The event is recognized as internal by using the measured time difference of two PMT 
signals compared to the estimated time of flight difference of the electrons.

The background can be classified in three groups: external one from incoming $\gamma$,
radon inside the tracking volume and internal radioactive contamination of the source.
All three were estimated from the NEMO~3 data with events of various topologies.
In particular, radon and internal 
$^{214}$Bi were measured with e$\gamma$$\alpha$ events. 
The e$\gamma$, e$\gamma$$\gamma$, and e$\gamma$$\gamma$$\gamma$ events
are used to measure the $^{208}$Tl activity requiring
the detection of the 2.615 MeV $\gamma$-ray typical of the $^{208}$Tl $\beta$-decay.
Single electron events are used to measure the foil contamination by $\beta$-emitters.
The external background is measured with the events with the detected incoming $\gamma$-ray producing 
an electron in the source foil. The external background is checked
with two-electron events
originating from pure copper and natural tellurium foils.
Measurements performed with an HPGe detector and radon detectors are used to verify the
results.

\section{NEMO~3 results}

\subsection{Measurement of $2\nu\beta\beta$ half-lives}

Measurements of the $2\nu\beta\beta$ decay half-lives were performed for 7 isotopes
available in NEMO~3 (see Table~\ref{tab:t12}). 
New preliminary results based on higher statistics than previously
are presented here 
for two of these isotopes: $^{48}Ca$ and $^{96}Zr$.
\begin{table}[hbt]
\caption{NEMO~3 results of half-life measurement.\label{tab:t12}}
\vspace{0.4cm}
\begin{center}
\begin{tabular}{ |c|c|c|c|c| }
\hline
Isotope& Mass (g) & Q$_{\beta\beta}$ (keV) & Signal/Background & T$_{1/2}$ [$10^{19}$ years]\\
\hline
$^{100}Mo$ &6914& 3034 & 40   & 0.711 $\pm$ 0.002 (stat) $\pm$ 0.054 (syst)~\cite{prl}\\
$^{82}Se$  &932 & 2995 & 4    & 9.6 $\pm$ 0.3 (stat) $\pm$ 1.0 (syst)~\cite{prl}\\
$^{116}Cd$ & 405& 2805 & 7.5  & 2.8 $\pm$ 0.1 (stat) $\pm$ 0.3 (syst)\\
$^{150}Nd$ &37.0& 3367 & 2.8  & $0.920 ^{+0.025}_{-0.022} $(stat) $\pm$ 0.062 (syst)\\
$^{96}Zr$  &9.4 & 3350 & 1.   & 2.3 $\pm$ 0.2 (stat) $\pm$ 0.3 (syst)\\
$^{48}Ca$  &7.0 & 4772 & 6.8  & $4.4 ^{+0.5}_{-0.4} $(stat) $\pm$ 0.4 (syst)\\
$^{130}Te$ &454 & 2529 & 0.25 & 76 $\pm$ 15 (stat) $\pm$ 8 (syst)\\
 \hline
\end{tabular}
\end{center}
\end{table}

\begin{figure}[htb]
\begin{center}
\includegraphics[width=0.41\textwidth,height=4.75cm]{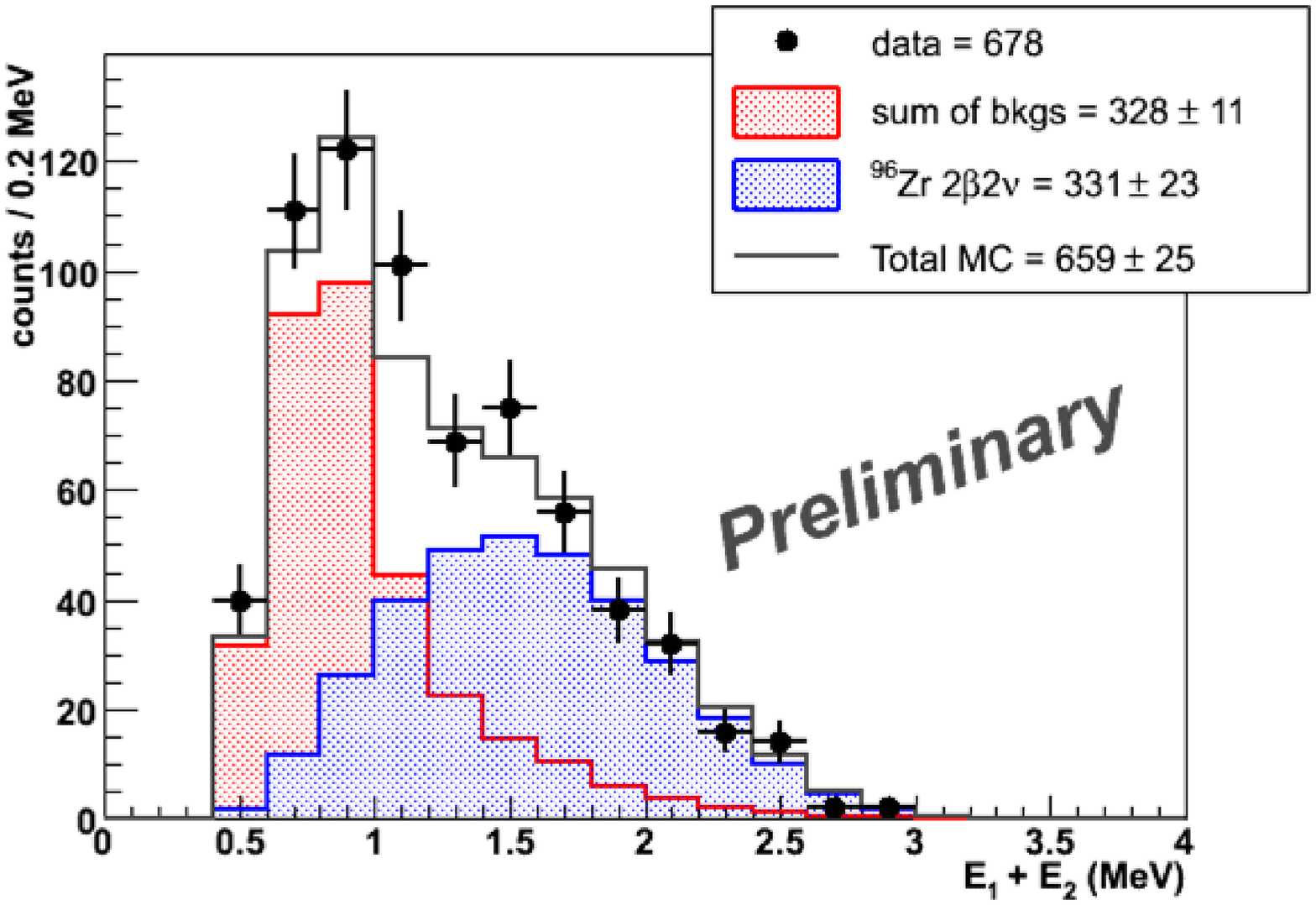}
\includegraphics[width=0.41\textwidth,height=4.75cm]{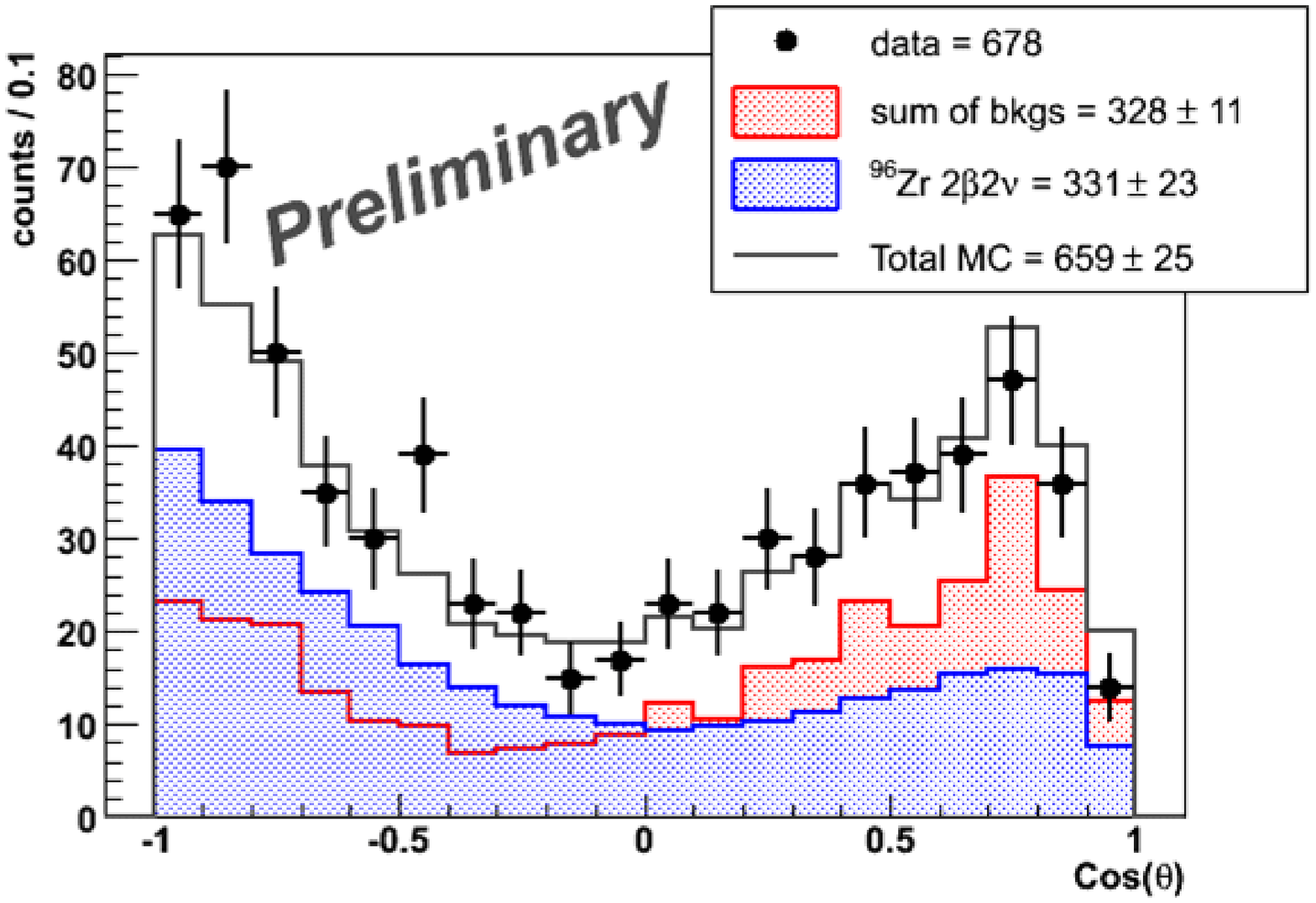}
\end{center}
\caption[Tl]{The energy sum and angular distributions
for two-electron events from $^{96}$Zr.}
\label{fig:bb_zr96}
\end{figure}

The measurement of the $^{96}Zr$ half-life was performed using the data collected within 
925 days. 
A total of 678 events were selected,  
with the expected 328 background events. The largest background
contribution is due to the internal $^{40}$K contamination of the sample. 
The distribution of the energy sum of two electrons and their angular distribution are shown 
in Fig.~\ref{fig:bb_zr96}, demonstrating good agreement of the data with the Monte Carlo simulation.
The $2\nu\beta\beta$ 
efficiency is estimated to be 7.6\%. The measured half-life is
$T_{1/2}^{2\nu}(^{96}Zr) = [2.3 \pm 0.2 (stat) \pm 0.3 (syst)]\cdot10^{19}y$.

\begin{figure}[htb]
\begin{center}
\includegraphics[width=0.41\textwidth,height=5.5cm]{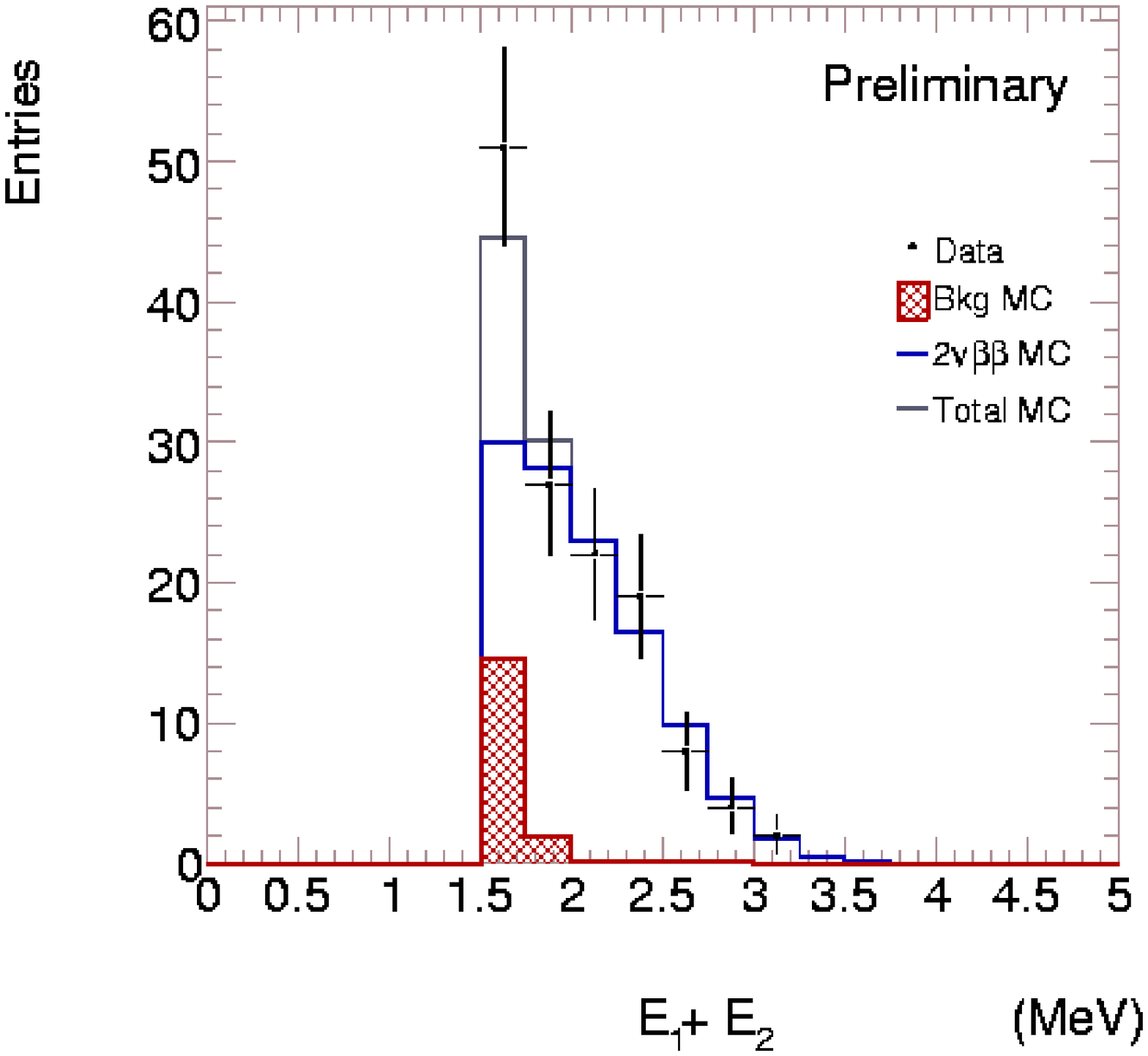}
\includegraphics[width=0.41\textwidth,height=5.5cm]{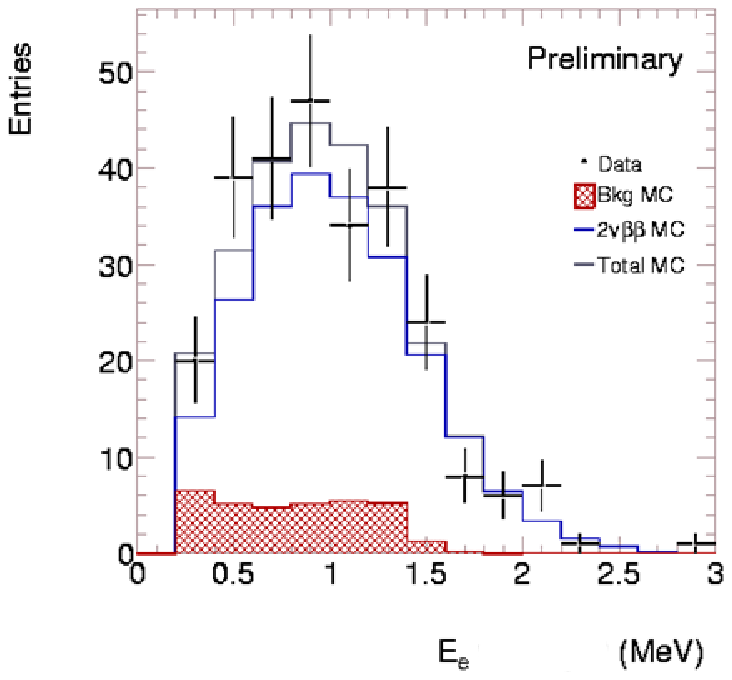}
\end{center}
\caption[Tl]{The energy sum and single-electron energy spectra  
for two-electron events from $^{48}$Ca.}
\label{fig:bb_ca48}
\end{figure}

The $^{48}$Ca sample used in NEMO~3 is known to be contaminated by 
$^{90}$Sr(T$_{1/2}$=28.79~y, $Q_{\beta}$=0.546~MeV). Its daughter 
$^{90}$Y (T$_{1/2}$=3.19~h, $Q_{\beta}$=2.282~MeV) is the major background source
in this case. An activity of $1699 \pm 3$ mBq/kg was 
measured for $^{90}$Y  using single-electron events.
Both $^{90}$Sr and $^{90}$Y are essentially pure $\beta^-$ emitters  
and imitate $\beta\beta$ events through M\"oller scattering. 
To supress this background contribution, events with  the energy sum  greater than 1.5 MeV 
and $cos(\Theta_{ee})<0$ are selected. 
Finally, with 943 days of data taking, there are a total of 133 two-electron events, 
with an evaluated residual background contribution of 17 events. 
Their two-electron energy sum distribution and single-electron energy spectrum
are presented in Fig.~\ref{fig:bb_ca48}.
The $2\nu\beta\beta$ efficiency is 3.3\%, and the measured half-life is 
$T_{1/2}^{2\nu}(^{48}Ca)=[4.4 ^{+0.5}_{-0.4} (stat) \pm 0.4 (syst)]\cdot10^{19}y$.

\subsection{Search for $0\nu\beta\beta$ decay}

In the case of the mass mechanism, the $0\nu\beta\beta$-decay signal is expected to be a peak 
in the energy sum distribution at the position of the transition energy $Q_{\beta\beta}$.
Since no excess is observed at the tail of the distribution  
for $^{96}$Zr, see Fig.~\ref{fig:bb_zr96} (left), nor for $^{48}$Ca, Fig.~\ref{fig:bb_ca48} (left),
limits are set on the neutrinoless double beta decay T$_{1/2}^{0\nu}$ using the CLs method~\cite{cls}.
A lower half-life limit is translated into an 
upper limit on the effective Majorana neutrino mass $\langle m_{\nu}\rangle$ (see Eq.~\ref{eq:nme}).
The following results are obtained using the NME values from~\cite{Kort}$^,$ \cite{Simkovic} for $^{96}$Zr
and  from~\cite{Caurier}  for $^{48}$Ca
\begin{tabbing}
T$_{1/2}^{0\nu}(^{96}Zr) > 8.6\cdot 10^{21} y$ (90\% C.L.) \hspace{1cm} \= $\langle m_{\nu}\rangle < $ 7.4 -- 20.1 eV \\
T$_{1/2}^{0\nu}(^{48}Ca) > 1.3\cdot 10^{22} y$ (90\% C.L.) \hspace{1cm} \= $\langle m_{\nu}\rangle < $ 29.7 eV.
\end{tabbing} 

\begin{figure}[htb]
\begin{center}
\includegraphics[width=0.41\textwidth,height=5.5cm]{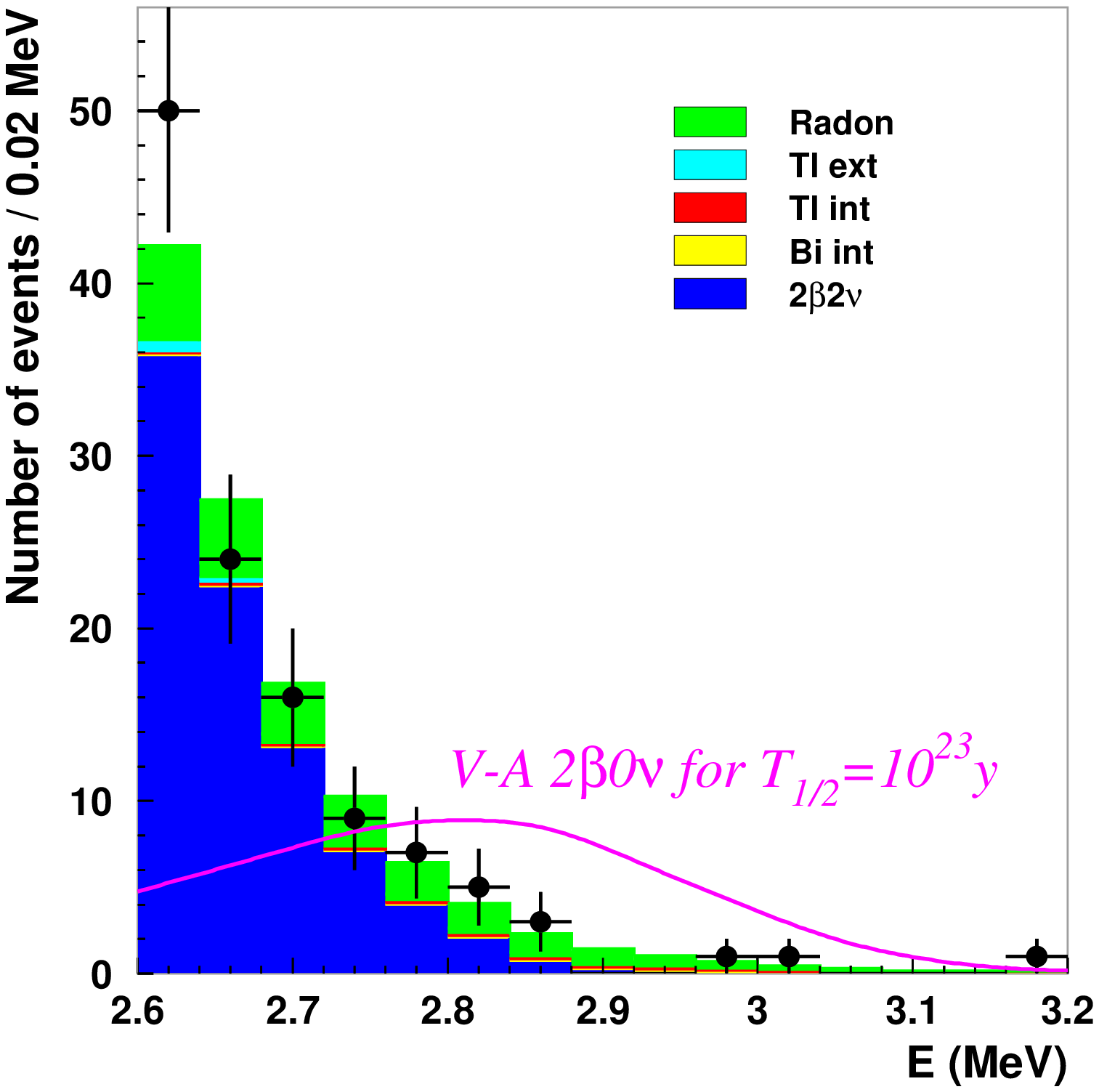}
\includegraphics[width=0.41\textwidth,height=5.5cm]{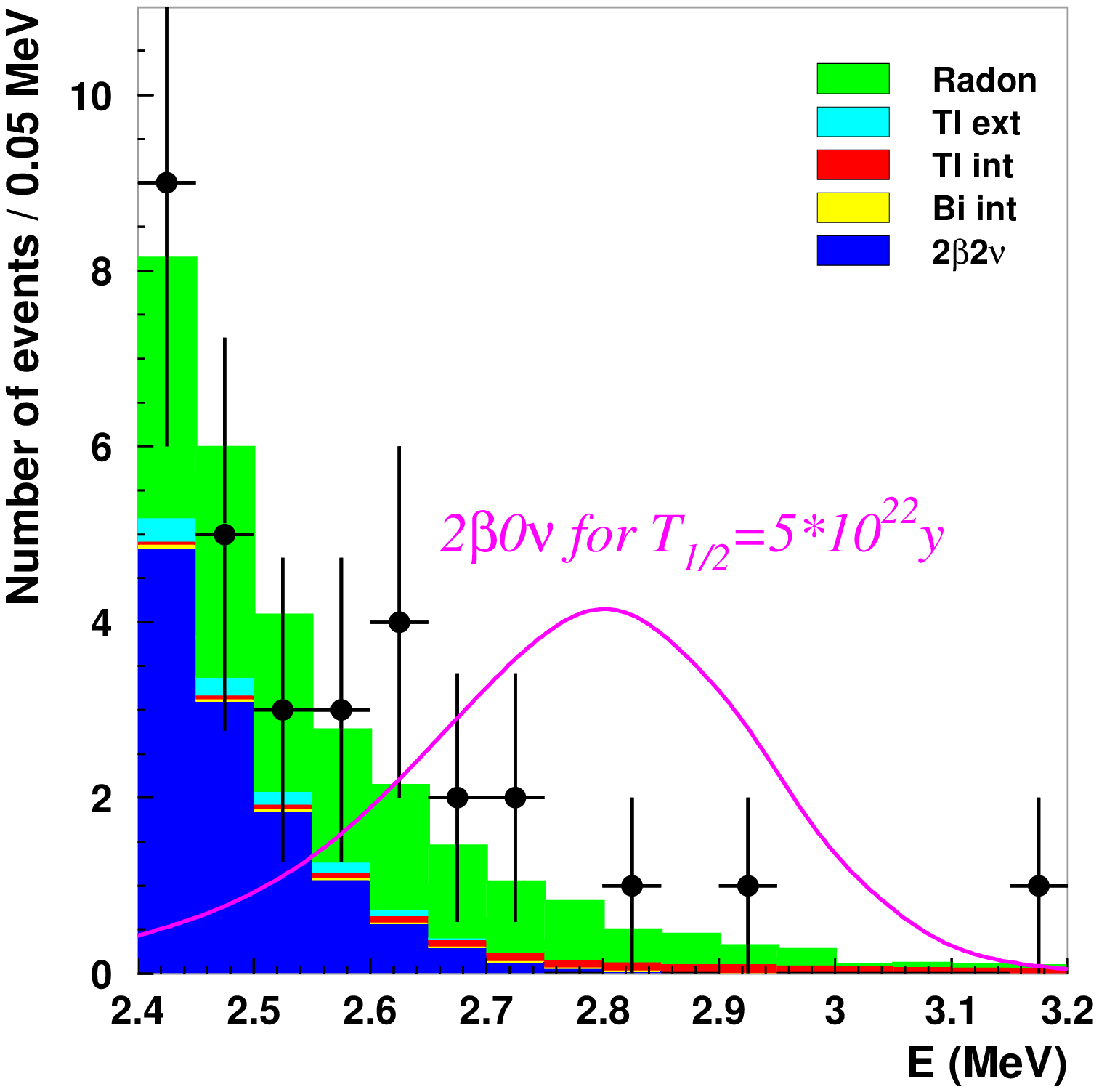}
\end{center}
\caption[Tl]{Distribution of the energy sum of two electrons 
for $^{100}$Mo (left) and  $^{82}$Se (right).}
\label{fig:0nu}
\end{figure}

The $0\nu\beta\beta$-decay search in NEMO~3 is most promising with $^{100}$Mo and $^{82}$Se 
because of the larger available isotope mass and high enough $Q_{\beta\beta} \sim$ 3 MeV.
The two-electron energy sum spectra obtained from the analysis of the data taken within 693 days
are shown in Fig.~\ref{fig:0nu}. For $^{100}$Mo there are 14 events observed in the $0\nu$
search window [2.78-3.20] MeV in good agreement with the expected background of 13.4 events.  
For $^{82}$Se there are 7 events found in the energy sum interval [2.62-3.20] MeV, compared to the expected 
background of 6.4 events. The limits on the T$_{1/2}^{0\nu}$ and the corresponding $\langle m_{\nu}\rangle$
limits calculated using the recent NME values~\cite{Kort}$^,$ \cite{Simkovic} are
\newline
T$_{1/2}^{0\nu}(^{100}Mo) > 5.8\cdot 10^{23} y$ (90\% C.L.) \hspace{0.9cm} $\langle m_{\nu}\rangle <$  0.61--1.26 eV \\
 T$_{1/2}^{0\nu}(^{82}Se) > 2.1\cdot 10^{23} y$ (90\% C.L.) \hspace{1.2cm} $\langle m_{\nu}\rangle <$   1.4 -- 2.2 eV.
\section{Summary}
The NEMO~3 experiment continues taking data to search for neutrinoless 
double beta decay  
with the target sensitivities
$\sim2\cdot10^{24}$ y for $^{100}Mo$ and $\sim 8\cdot 10^{23}$ y for $^{82}Se$
 in terms of  90\% C.L. limit on T$_{1/2}^{0\nu}$.
No evidence for a signal was observed after 693 days of effective data collection.
The 90\% C.L. limits were obtained:
T$_{1/2}^{0\nu}(^{100}Mo) > 5.8\cdot 10^{23} y$ and 
T$_{1/2}^{0\nu}(^{82}Se) > 2.1\cdot 10^{23} y$.
 
 The half-lives of 
$2\nu\beta\beta$-decay of  $^{100}$Mo,$^{82}$Se,$^{116}$Cd,$^{130}$Te,$^{150}$Nd,
$^{96}$Zr and $^{48}$Ca were measured in NEMO~3.
New preliminary results were obtained for 
two of these isotopes:
$T_{1/2}^{2\nu}(^{96}Zr)=[2.3 \pm 0.2 (stat) \pm 0.3 (syst)]\cdot10^{19}y$,
$T_{1/2}^{2\nu}(^{48}Ca)=[4.4 ^{+0.5}_{-0.4} (stat) \pm 0.4 (syst)]\cdot10^{19}y$.


\end{document}